\documentclass[
twocolumn,
showpacs,
superscriptaddress,
nofootinbib,
amssymb,
a4paper
]{revtex4-1}
\usepackage[dvipdfmx]{graphicx}
\usepackage{ulem}
\usepackage{color}
\usepackage{bm}
\usepackage{braket}
\usepackage{amsmath,amssymb}
\usepackage{comment}
\usepackage{lineno}
\usepackage{mathtools}

\def\({\left(}
\def\){\right)} 
\def\[{\left[}
\def\]{\right]} 

\newcommand{\slh}[1]{ {#1} \hspace{-0.65em}/} 
\newcommand{\Tr}{\mathrm{Tr}}

\newcommand{\Det}{\mathrm{Det}}

\newcommand{\UA}{${\rm U}(1)_A$~}

\begin{document}
%
%
%
\title{
  Different scenarios of dynamical chiral symmetry breaking 
  in the interacting instanton liquid model via flavor symmetry breaking
}
%
\author{
  Yamato Suda
}
\author{
  Daisuke Jido
}
\affiliation{
  Department of Physics, 
  Institute of Science Tokyo,
  2-12-1 Ookayama, Meguro, Tokyo 152-8551, 
  Japan
}
%
%
\date{\today}

\begin{abstract}
We investigate a type of dynamical chiral symmetry breaking (D$\chi$SB) 
for various current quark masses using the interacting instanton liquid model. 
The type of D$\chi$SB is classified based on the sign of the second derivative 
of the free energy density with respect to the quark condensate at the origin. 
We perform numerical simulations of the interacting instanton liquid model 
with the flavor SU(2) symmetric and (2+1)-flavor quarks. We find that the 
curvature is negative in the SU(2) case. This means the ordinary type of D$\chi$SB. 
In contrast, in the (2+1)-flavor case, a positive curvature is observed 
when the strange quark mass is as small as those of the up and down quarks.
This suggests that the anomaly-driven type of D$\chi$SB can occur under 
the approximate flavor SU(3) symmetry. As the strange quark mass increases, 
the curvature gradually decreases and becomes negative when the strange quark 
mass is approximately three times larger than those of the light quarks. 
This difference can be understood in terms of the 't Hooft vertex which 
induces a six-quark interaction in the $N_f=3$ case and does a four-quark 
interaction in the $N_f=2$ case. Our results might indicate that the ratio 
between the strange and light quark masses plays a crucial role in understanding 
the microscopic relationship between D$\chi$SB and the anomaly effect.
\end{abstract}
%
%
\maketitle
%
%

\section  {Introduction}
\label{sec:Introduction}

Understanding the nature of dynamical chiral symmetry breaking (D$\chi$SB) 
in the vacuum of strong interaction is one of the most important subjects 
in hadron physics~\cite{Shuryak:2021}. The fundamental theory of strong 
interaction, quantum chromodynamics (QCD), is invariant under the chiral 
transformation at the Lagrangian level in the chiral limit. Due to the 
non-perturbative dynamics of quarks and gluons, quark and anti-quark are
condensed in the vacuum, leading to the dynamical breaking of chiral symmetry as
\begin{eqnarray}
  {\rm SU}(3)_L \times {\rm SU}(3)_R \times {\rm U}(1)_A \to {\rm SU}(3)_V.
\end{eqnarray}
The \UA symmetry is also broken by a quantum anomaly initially identified 
by Adler, Bell and Jackiw~\cite{Adler:1969,BellJackiw:1969}. Due to this 
anomaly, called the axial anomaly, the singlet $\eta'$ meson acquires 
relatively large mass than the other Nambu--Goldstone 
bosons~\cite{Kawarabayashi:1980,Kawarabayashi:1981,Bass:2019}.
D$\chi$SB is responsible not only for the most of the hadron mass generation, 
but also for the non-trivial vacuum structure of strong interaction. 

The connection between the axial anomaly and D$\chi$SB 
in the vacuum of strong interaction has been discussed 
for a long time~\cite{Shuryak:1982_1, Shuryak:1982_2, Schafer:1998,Zahed:2021,Braghin:2024}. 
As an earlier work, 't Hooft explained the chiral anomaly 
is induced by a vacuum with a single instanton~\cite{tHooft:1976_b}, 
and afterward the explanation of D$\chi$SB had been provided 
as the formation of non-zero density of states around 
zero eigenvalue of Dirac operator due to the multi-instanton vacuum 
configuration~\cite{Shuryak:1988,Nowak:1989a,Nowak:1989b,Nowak:1998,Leutwyler:1992}, 
which is also known as the Banks--Casher 
relation~\cite{Banks:1980,Smilga:1993,Verbaarschot:1993,Spitzenberg:2002}. 

Turning our attention to the chiral effective theories such as the chiral 
perturbation theory, the linear sigma model and the Nambu--Jona-Lasinio (NJL) 
model, those are based on the symmetry structure and its breaking pattern to 
clarify the underlying physics~\cite{Weinberg:1979,Gasser:1985,GellMann:1960,Nambu:1961}. 
Their successes are often regarded as being independent of the details 
of how chiral symmetry is broken in the vacuum, in other words, independent 
of the microscopic mechanism behind it. However, it remains an open and 
interesting question what kind of mechanism is predominantly responsible 
for D$\chi$SB and whether it leaves observable remnants in hadron properties.

In the previous study~\cite{Kono:2021}, the authors investigated how the 
mechanism predominantly responsible for D$\chi$SB in vacuum---either the 
four-quark interaction or the anomalous \UA dynamics---can affect not only 
the vacuum structure but also the hadron spectrum. Using the NJL model with 
the determinant-type Kobayashi--Maskawa--'t Hooft (KMT) 
interaction~\cite{Kobayashi:1970,Kobayashi:1971,tHooft:1976_a,tHooft:1976_b,Rosenzweig:1980,Vecchia:1980,Witten:1980},
they classified the types of D$\chi$SB based on the relative strength of the 
four-quark coupling $g_S$ to its critical value $g_S^{\rm crit}$ with the 
anomaly-induced coupling $g_D$. When $g_S > g_S^{\rm crit}$, D$\chi$SB occurs 
even without the anomaly effect, and this situation is referred to as the 
ordinary type. In contrast, when $g_S<g_S^{\rm crit}$, D$\chi$SB can still 
occur if the anomaly term is sufficiently strong, and this case is defined 
as the anomaly-driven type. These are two types of 
interactions---four-quark and anomaly---that are 
predominantly responsible for D$\chi$SB. Furthermore, their 
analysis revealed that this classification leaves observable footprints 
in the hadron spectrum: in the ordinary type, the $\sigma$ meson, which is 
introduced as the chiral partner of the pion, obtains a heavier mass 
$m_\sigma > 800~{\rm MeV}$, whereas in the anomaly-driven type, it becomes 
lighter with $m_\sigma < 800~{\rm MeV}$. This suggests that the $\sigma$ 
meson mass may serve as a probe of the underlying dynamics driving D$\chi$SB.

In Ref.~\cite{Suda:2024}, we investigated whether such different types of 
D$\chi$SB can also emerge in models other than those considered in 
Ref.~\cite{Kono:2021}. To generalize the discussion in Ref.~\cite{Kono:2021} 
beyond the scope of specific models, we first defined a criterion to 
identify the type of D$\chi$SB based on the sign of the curvature, 
which is defined as the second derivative of the energy density of 
the vacuum with respect to the quark condensate at the origin~\cite{Suda:2024}. 
When this curvature is negative or positive, we classify the emergent 
D$\chi$SB as the ordinary type or the anomaly-driven type, respectively. 
This provides a generalized criterion compared to the one based on the 
couplings $g_S$ and $g_D$. By calculating the curvature of the effective 
potential by using the NJL model with $g_S>g_S^{\rm crit}$ regardless 
of $g_D$, we find a negative curvature. In contrast, for $g_S<g_S^{\rm crit}$ 
with a sufficiently large $g_D$, the curvature at the origin becomes positive.

In our previous study~\cite{Suda:2024}, we performed numerical simulations 
of the interacting instanton liquid model (IILM) and evaluated the curvature 
both in IILM with the flavor SU(3) symmetric quarks and in the quenched calculation. 
The previous work concluded that IILM with the flavor SU(3) symmetric quarks 
exhibits the anomaly-driven type of D$\chi$SB, while the quenched calculation 
shows the ordinary type of D$\chi$SB~\cite{Suda:2024}.

This difference can be understood from the perspective of quark-instanton 
interactions mediated by the so-called 't Hooft 
vertex~\cite{tHooft:1976_b,CDG:1978,CDG:1979,Shifman:1980}. In the quenched 
calculation, the quark determinant is omitted from the partition function. 
This removes the dynamical interaction between quarks and instantons. 
As a result, the multi-fermion vertex induced by instantons that involves 
$2N_f$ quark legs and brings the effects of the \UA anomaly does not arise 
in the quenched calculation. This leads to a suppression of anomaly effects.
Thus, the D$\chi$SB mechanism is found to be of the ordinary type.

In contrast, IILM with the flavor SU(3) symmetric quarks includes the full 
quark determinant that provides the dynamical quark-instanton interaction. 
The resulting 't Hooft vertex in this case is a six-quark interaction.
This corresponds to the KMT term in chiral effective theories. Since this 
interaction originates from the \UA anomaly, the effect of the axial anomaly 
is included in this calculation. Therefore, in the flavor SU(3) symmetric IILM, 
the anomaly-driven type of D$\chi$SB is naturally understood. From this, we 
anticipate that the presence of three quarks with the flavor SU(3) symmetry 
may play an important role in inducing the anomaly-driven type of D$\chi$SB. 

In nature, the $u$ and $d$ quarks have nearly equal masses, $m_u \approx m_d$, 
while the $s$ quark is significantly heavier, $m_s \gg m_u,m_d$~\cite{PDG:2024}. 
Therefore, the flavor symmetry that more closely reflects the physical situation 
is SU(2) and it is nontrivial to extrapolate our conclusions in the flavor SU(3) 
symmetric and quenched calculations to the physical case, where the flavor SU(3) 
symmetry is slightly broken due to the mass difference among the $u,d$ and $s$ 
quarks. This implies that further investigation is necessary to understand how 
the anomaly-driven type of D$\chi$SB is physically emerged. 

In this study, we investigate the dependence of the type of D$\chi$SB on the 
number of quark flavors using IILM with the flavor SU(2) symmetric and 
(2+1)-flavor quarks. The aim of this study is to clarify how the type of 
D$\chi$SB changes as we move from the flavor SU(3) symmetric system to the 
flavor SU(2) symmetric one by increasing the strange quark mass. This can be 
one of the clues to understanding the nature of D$\chi$SB occurring in the 
vacuum of strong interaction.

This paper is organized as follows. In Sect.~\ref{sec:methodology}, 
we demonstrate the formulation of IILM and computational details used 
in this study. In Sect.~\ref{sec:results}, we show the results of the 
flavor SU(2) and (2+1)-flavor calculations with various quark masses.
In Sect.~\ref{sec:conclusion}, we summarize this paper.

\section  {Methodology}
\label{sec:methodology}

In this section, we will explain 
the model used in this study 
and the computational procedures 
for measurements relevant to our evaluations.
In addition, we will describe how to 
extract the curvature from those quantities 
in the numerical simulation of IILM. 

\subsection  {Partition function}
\label{sec:PartitionFunction}

We begin with the partition function of IILM 
in the four-dimensional Euclidean space-time given by Ref.~\cite{Schafer:1996}:  
\begin{eqnarray} 
  Z 
  &=& \frac{1}{N_+ ! N_- !} \int \prod_{i=1}^{N_+ + N_-} 
  [d \Omega_i f(\rho_i)] \exp (-S_{\rm int}) \nonumber \\ 
  &\ & \quad \times \prod_{f} \Det (\slh{D} + m_f), \label{eq:IILM_Z}
\end{eqnarray}
where the number of instantons and anti-instantons 
are denoted as $N_\pm$. The path integral measure is represented 
by $d\Omega_i = dU_i d^4z_i d\rho_i$, 
which consists of the collective coordinates, 
such as color orientation $U$, position $z$, and size $\rho$
of an instanton or anti-instanton 
labeled by $i$. 
The semiclassical instanton amplitude 
$f(\rho)$ as a function of the instanton size $\rho$
depends on the number of colors $N_c=3$ and 
of flavors $N_f$ via the $\beta$ functions.
This is originally 
calculated by 't Hooft and 
extended to SU($N_c$) group 
by Bernard~\cite{tHooft:1976_b,Bernard:1979}.
In the calculation of $f(\rho)$, 
a scale parameter $\Lambda$ 
is introduced~\cite{Dyakonov:1986}.
Its explicit form is given in Appendix.~\ref{app:semiclassicalamp}.
All quantities in this study are computed 
in units of the scale parameter $\Lambda$.

The action $S_{\rm int}$ accounts for 
the interaction between instantons. 
According to Ref.~\cite{Schafer:1996}, 
we approximate it by a sum of a purely two-body interaction 
$S_{\rm int} = (1/2) \sum_{l \neq m} S^{(2)}_{\rm int}(l,m)$ 
for all possible combinations of instantons and anti-instantons.
The quark determinant $\Det (\slh{D}+m_f)$ 
arises from the path integral of degrees of freedom 
of quarks with a current mass $m_f$. 
This introduces a nonlocal interaction 
among instantons. The details for these interactions 
are described in Sect.~\ref{sec:Interactions}.

The dynamics of instantons and anti-instantons 
described by the partition function in Eq.~(\ref{eq:IILM_Z})
can be simulated by using 
standard Monte Carlo techniques, 
such as the Metropolis 
algorithm and the HMC algorithm, originally 
developed for simulations of statistical mechanics 
of atoms or molecules~\cite{Metropolis:1953} 
and lattice field theory~\cite{Duane:1987}. 
In the present case, 
we fix a total number of instantons and anti-instantons, 
$N\equiv N_++N_-=16+16=32$. 
We then perform a simulation of 
IILM according to Eq.~(\ref{eq:IILM_Z}) 
at a given instanton density, 
$n \equiv N/V$, by adjusting a four-volume of the system $V$.  
To investigate the instanton density dependence 
of measurements, we vary the four-volume $V$ 
while keeping $N$ fixed.  

In this simulation, the remaining input parameters
are current quark masses, $m_f=(m_u,m_d,m_s)$.
In practice, the current quark masses 
are specified in units of $\Lambda$ 
at the beginning of the simulation.
Once $\Lambda$ is determined for given quark masses $m_f$, 
the quantities given in units of $\Lambda$ 
can be converted into physical units such as 
MeV and fm.

We determine the value of
the scale parameter $\Lambda$ 
such that the free energy density
has its minimum at an instanton 
density of $n=1~\rm{\rm fm}^{-4}$, 
following Ref.~\cite{Schafer:1996}.
The definition of the free energy density 
is given in Sect.~\ref{sec:Measurements}.

\subsection  {Interactions}
\label{sec:Interactions}

The two-body interaction 
$S^{(2)}_{\rm int}(l,m) = S[A_\mu (l,m)] - 2 S_0$ 
is calculated by inserting the instanton-anti-instanton 
gauge potential. Here, the single instanton action is denoted as 
$S_0 = 8 \pi^2 / g^2$.
For this potential, 
we use the streamline configuration~\cite{Verbaarschot:1991}.
In practice, we introduce a phenomenological short 
range repulsive core in both the instanton-instanton and 
anti-instanton-anti-instanton interactions.
This includes a parameter $A$ that controls the strength of the core. 
We use $A=128$ that is tuned so as to make the ensemble 
not as dilute as that suggested by phenomenology~\cite{Schafer:1996}.
The explicit expression of $S_{\rm int}$ 
is given in Appendix.~\ref{app:interaction}.

The quark determinant for each quark flavor $f$ 
can be written as a product of determinants over high and low
momentum parts~\cite{Dyakonov:1986}:
\begin{eqnarray}
  \Det (\slh{D} + m_f) 
  &\equiv& \Det_{\rm high} (\slh{D} + m_f) \Det_{\rm low} (\slh{D} + m_f) \nonumber \\ 
  &\equiv& \left( \prod_{i=1}^N 1.34 \rho_i \right) \Det_{\rm I,\bar I}(-i T + m_f \bm{1}),
  \label{eq:Det}
\end{eqnarray}
where the first factor, the high momentum part, 
is given as a product of contributions from the background field 
of individual instantons, whereas the low momentum part 
responsible for the quark zero-mode in the 
instanton and anti-instanton backgrounds is calculated exactly. 
Here, $T$ represents the overlap matrix with a size of 
$N \times N$. This matrix is spanned by 
the quark zero-mode wave functions associated with the instanton or 
the anti-instantons. Thus, the determinant $\Det_{\rm I,\bar I}$
is understood as an operation carried out over 
the space spanned by the quark zero-modes of size $N \times N$.
The explicit expression of $T$ is also 
given in Appendix.~\ref{app:interaction}.

For the flavor SU(2) symmetric case,
we use $N_f=2$ in the calculation of $f(\rho)$, 
while for the case of the (2+1)-flavor quarks, 
we use $N_f=3$ for that. 
In addition, we treat the quark determinant part
of Eq.~(\ref{eq:IILM_Z}) as
\begin{align}
  \displaystyle\prod_{f} \Det (\slh{D} + m_f)
  = \[ \Det (\slh{D} + m_q)\]^{N_f},
  \label{eq:FlavorSym}
\end{align}
for the flavor symmetric cases such as 
the flavor SU(3) and SU(2) calculations, and 
\begin{eqnarray}
  \displaystyle\prod_{f} \Det (\slh{D} + m_f)
  = \displaystyle\prod_{f=q,q,s} \Det (\slh{D} + m_f) 
  \label{eq:FlavorAsym}
\end{eqnarray}
for the (2+1)-flavor case. 
In this work, we assume that the masses of the $u$ and $d$ quarks 
are equal and denote them as $m_q \equiv m_u = m_d$.
The $s$ quark mass is denoted as $m_s$.

This quark determinant for a single flavor originates from 
the path integral over the fermionic degrees of freedom for 
that flavor~\cite{Schafer:1998}. Thus, the single-flavor determinant 
corresponds to a two-quark vertex in the effective Lagrangian of 
the instanton model discussed in Ref.~\cite{Shifman:1980}. 
This implies that the determinant structure in Eq.~(\ref{eq:FlavorSym}) 
can be interpreted as including up to four-quark interactions
for $N_f=2$. In contrast, the determinant form in Eq.~(\ref{eq:FlavorAsym}) 
may correspond to a six-quark effective interaction that introduces 
the effect of the axial anomaly to the system for $N_f=3$. This is 
the significant difference between the flavor SU(2) symmetric case 
and the (2+1)-flavor case in this model.

\subsection  {Measurements}
\label{sec:Measurements}

The free energy density $F$ 
is given by logarithm of the partition function 
in Eq.~(\ref{eq:IILM_Z}), 
divided by the volume $V$:
\begin{eqnarray}
  F = - \frac{1}{V} \ln Z.
\end{eqnarray}
In the following, 
we simply refer to this as the free energy.
The value of $F$ is calculated 
at each instanton density $n$ 
for fixed quark masses, $(m_q, m_q)$ 
for the flavor SU(2) symmetric calculations 
and $(m_q,m_q,m_s)$ for the (2+1)-flavor ones.
To compute the free energy,
we use the thermodynamics integration method, which was first applied
to this framework by Sch\"afer and Shuryak in Ref.~\cite{Schafer:1996}.

The quark condensate for a single flavor $f$ of mass $m_f$ 
is evaluated as the expectation value of 
the traced quark propagator at the same coordinate:
\begin{eqnarray}
  \braket{\bar \psi_f \psi_f} 
  &\approx& - \lim_{y \to x} \frac{1}{Z} 
  \int D \Omega\ e^{-S_{\rm int}} 
  \prod_{f'} \Det (\slh{D}+m_{f'}) \nonumber \\ 
  &\ & \qquad \times \Tr \[S^{\rm ZM}(x,y;m_f)\], \label{eq:qq_def}
\end{eqnarray}
where we write $D\Omega \equiv \prod_i^N[d\Omega_i f(\rho_i)]$ in short, 
and we approximate 
the full quark propagator $S(x,y;m_f)$ 
by the zero-mode propagator $S^{\rm ZM}(x,y;m_f)$~\cite{Dyakonov:1986}. 
The trace $\Tr$ is performed over the Dirac and color indices. 
The value of $\braket{\bar \psi_f \psi_f}$ is also calculated 
at each instanton density $n$ once we specify 
the quark masses. 
The explicit form of the zero-mode propagator 
is given by 
\begin{eqnarray}
  &\ & S^{\rm ZM}(x,y;m_f) \nonumber \\ 
  &\ & \qquad = \sum_{I,J}\[\psi_{0,I}(x)\[(-iT + m_q)^{-1}\]_{I,J} \psi^\dag_{0,J}(y)\]. \label{eq:ZM_prop}
\end{eqnarray}
Here, $\psi_{0,I}(x)$ is the quark zero-mode wave 
function in the instanton background 
and $T$ is the overlap matrix defined in Eq.~(\ref{eq:Det}). 
For further details, see Refs.~\cite{Schafer:1998,Suda:2024}. 
We simply write the $u$ and $d$ quark condensates
by $\braket{\bar qq} \equiv \braket{\bar \psi_u \psi_u} = \braket{\bar \psi_d \psi_d}$, 
and refer this as light quark condensate, 
whereas we explicitly denote the flavor 
for the $s$ quark condensate
as $\braket{\bar ss} \equiv \braket{\bar \psi_s \psi_s}$.

By computing the free energy and the quark condensate 
at each instanton density, 
for given quark masses we obtain the set of measurements 
as a function of the instanton density 
as $(F_j, \braket{\bar qq}_j)$ for the flavor SU(2) symmetric 
and $(F_j,\braket{\bar qq}_j, \braket{\bar ss}_j)$
for the (2+1)-flavor calculations, where index $j$ labels 
the instanton density. 

To evaluate the curvature for given quark masses $m_f$, 
we perform the polynomial regression 
on the dataset $(F_j, \braket{\bar qq}_j)$ 
for both cases of the flavor SU(2) symmetric and the (2+1)-flavor.
The polynomial model that to be fitted is given by 
\begin{eqnarray}
  F
  = C_0 
  + C_1 \braket{\bar{q}q} 
  + \frac{1}{2} C_2 \braket{\bar{q}q}^2 
  + \cdots 
  + \frac{1}{k!} C_k \braket{\bar{q}q}^k, \quad
  \label{eq:polynomial_model}
\end{eqnarray}
with the polynomial order $k$. 
We vary the order $k$ 
from 1 to 3 in this study 
to investigate systematic variation. 
We will show only the results of $k=3$ 
in Sect.~\ref{sec:results} because we 
find no qualitative difference 
between the results of $k=2$ and $3$. 
In the following, we simply refer to 
the coefficient $C_2$ as curvature.

As we have mentioned in Sect.~\ref{sec:Introduction},
one of the key characteristics to classify the type
of D$\chi$SB is the sign of this curvature.
We define the anomaly-driven type of chiral symmetry breaking 
as occurring in the vacuum when the curvature is positive, 
and the ordinary type of chiral symmetry breaking as occurring 
in the vacuum when the curvature is negative. 

\subsection  {Computational setups}
\label{sec:ComputationalSetups}

We perform the numerical simulations 
of IILM with the flavor SU(2) symmetric and (2+1)-flavor quarks. 
For the flavor SU(2) symmetric calculation,
we prepare six sets of simulations at different quark masses 
with setting the $u$ and $d$ quarks to have equal mass $m_q$. 
For the (2+1)-flavor calculation, 
we prepare 37 sets of simulations at different quark masses 
where the $u$ and $d$ quarks have equal mass $m_q$
while the $s$ quark is heavier; $m_q<m_s$.
The values of $m_q$ and $m_s$ in units of $\Lambda$ 
are summarized in Table.~\ref{tabel:configuration_Nf2} 
for $N_f=2$ and Table.~\ref{tabel:configuration_Nf21} 
for $N_f=2+1$.

In performing those simulations, 
we use the standard Metropolis algorithm 
for updating the color orientation $U_i$, 
and we use the HMC algorithm for updating the 
size $\rho_i$ and position $z_i$ of an $i$th 
instanton~\cite{Binder:2010,Randau:2014,Hanada:2018}. 
In scanning the instanton density dependence 
of the measurements, we calculate
with 72 different instanton densities.
One simulation at a given instanton density 
and quark masses consists of 
$5000$ configurations
that are generated after $1000$ initial sweeps.

\begin{table}
  \caption{
    \label{tabel:configuration_Nf2}  
    Values of $m_q$ in units of $\Lambda$ for the simulations of 
    IILM with the flavor SU(2) symmetric quarks.
    Values of the scale parameter $\Lambda$ are determined 
    following the procedure in the text. 
    Values of $m_q$ in units of MeV are calculated 
    by using these values of $\Lambda$.
  }
    \begin{tabular*}{86mm}{l@{\extracolsep{\fill}}cccccc} \hline \hline 
      Sets & B1 & B2 & B3 & B4 & B5 & B6 \\ \hline
      $m_q~(\Lambda)$ & $0.05$ & $0.08$ & $0.10$ & $0.15$ & $0.18$ & $0.20$ \\
      $\Lambda~({\rm MeV})$ & $331$ & $322$ & $325$ & $320$ & $316$ & $316$ \\
      $m_q~({\rm MeV})$ & $17$ & $26$ & $32$ & $48$ & $57$ & $63$ \\ 
      \hline \hline
    \end{tabular*}
\end{table}

\begin{table*}
  \caption{
  \label{tabel:configuration_Nf21}
  Quark masses $m_f=(m_q,m_q,m_s)$ in units of $\Lambda$ for the simulations 
  of IILM with the (2+1)-flavor quarks. Values of $\Lambda$ and $(m_q,m_s)$ 
  in units of MeV are calculated in the same way as in Table.~\ref{tabel:configuration_Nf2}.
  }
  \begin{tabular*}{0.99\textwidth}{l@{\extracolsep{\fill}}lllllll} \hline \hline 
    Sets & C1 & C2 & C3 & C4 & C5 & C6 & \\ \hline
    $(m_q,m_s)~(\Lambda)$ & $(0.08, 0.10)$ & $(0.08, 0.15)$ & $(0.08, 0.30)$ & $(0.08, 0.60)$ & $(0.08, 0.90)$ & $(0.08, 1.2)$ \\
    $\Lambda~(\mathrm{MeV})$ & $394$ & $379$ & $364$ & $336$ & $326$ & $314$ \\
    $(m_q,m_s)~(\mathrm{MeV})$ & $(31,39)$ & $(30,57)$ & $(29,109)$ & $(27,202)$ & $(26,293)$ & $(25,377)$ \\
    \hline \hline 
    Sets & D1 & D2 & D3 & D4 & D5 & D6 & D7 \\ \hline
    $(m_q,m_s)~(\Lambda)$ & $(0.10, 0.15)$ & $(0.10, 0.30)$ & $(0.10, 0.45)$ & $(0.10, 0.60)$ & $(0.10, 0.75)$ & $(0.10, 0.90)$ & $(0.10, 1.2)$ \\
    $\Lambda~(\mathrm{MeV})$ & $372$ & $358$ & $349$ & $338$ & $331$ & $322$ & $319$ \\
    $(m_q,,m_s)~(\mathrm{MeV})$ & $(37,56)$ & $(36,108)$ & $(35,157)$ & $(34,203)$ & $(33,249)$ & $(32,290)$ & $(32,383)$ \\
    \hline \hline 
    Sets & E1 & E2 & E3 & E4 & E5 & E6 \\ \hline
    $(m_q,m_s)~(\Lambda)$ & $(0.15, 0.20)$ & $(0.15, 0.30)$ & $(0.15, 0.45)$ & $(0.15, 0.60)$ & $(0.15, 0.90)$ & $(0.15, 1.2)$ \\
    $\Lambda~(\mathrm{MeV})$ & $365$ & $353$ & $335$ & $329$ & $322$ & $310$ \\
    $(m_q,m_s)~(\mathrm{MeV})$ & $(55,73)$ & $(53,106)$ & $(50,151)$ & $(49,197)$ & $(48,290)$ & $(47,372)$ \\
    \hline \hline 
    Sets & F1 & F2 & F3 & F4 & F5 & F6 \\ \hline
    $(m_q,m_s)~(\Lambda)$ & $(0.20, 0.25)$ & $(0.20, 0.30)$ & $(0.20, 0.45)$ & $(0.20, 0.60)$ & $(0.20, 0.90)$ & $(0.20, 1.2)$ \\
    $\Lambda~(\mathrm{MeV})$ & $ 342 $ & $ 340 $ & $ 331 $ & $ 322 $ & $ 311 $ & $ 303 $ \\
    $(m_q,m_s)~(\mathrm{MeV})$ & $ (68,86) $ & $ (68,102) $ & $ (66,149) $ & $ (64,193) $ & $ (62,280) $ & $ (61,364) $ \\
    \hline \hline 
    Sets & G1 & G2 & G3 & G4 & G5 & G6 \\ \hline
    $(m_q,m_s)~(\Lambda)$ & $(0.25, 0.28)$ & $(0.25, 0.30)$ & $(0.25, 0.45)$ & $(0.25, 0.60)$ & $(0.25, 0.90)$ & $(0.25, 1.2)$ \\
    $\Lambda~(\mathrm{MeV})$ & $337$ & $332$ & $323$ & $320$ & $307$ & $302$ \\
    $(m_q,m_s)~(\mathrm{MeV})$ & $(84, 94)$ & $(83,99)$ & $(81,145)$ & $(80,192)$ & $(77,276)$ & $(76,363)$ \\
    \hline \hline 
    Sets & H1 & H2 & H3 & H4 & H5 & H6 \\ \hline
    $(m_q,m_s)~(\Lambda)$  & $(0.30, 0.45)$ & $(0.30, 0.60)$ & $(0.30, 0.75)$ & $(0.30, 0.90)$ & $(0.30, 1.0)$ & $(0.30, 1.2)$ \\
    $\Lambda~(\mathrm{MeV})$ & $319$ & $313$ & $308$ & $301$ & $300$ & $296$ \\
    $(m_q,m_s)~(\mathrm{MeV})$ & $(96, 143)$ & $(94,188)$ & $(92,231)$ & $(90,271)$ & $(90,300)$ & $(89,355)$ \\
    \hline \hline 
  \end{tabular*}
\end{table*}

\section  {Results}
\label{sec:results}

We focus on the curvature $C_2$ obtained from 
the $\braket{\bar{q}q}$ dependence of $F$.
In Sect.~\ref{subsec:flavor-sym-res}, 
we will present the results for the flavor SU(2) 
symmetric case and compare 
it to the flavor SU(3) symmetric case
obtained in our previous study~\cite{Suda:2024}. 
In Sect.~\ref{subsec:2+1-flavor-res}, 
we will show the results for the (2+1)-flavor calculations. 
Finally, in Sect.~\ref{subsec:transition-res}, 
we will examine the difference between 
the flavor SU(3) and SU(2) symmetric cases
by analyzing the (2+1)-flavor results.

\subsection {The flavor SU(2) symmetric calculation}
\label{subsec:flavor-sym-res}

\begin{figure}
  \includegraphics[width=0.46\textwidth]{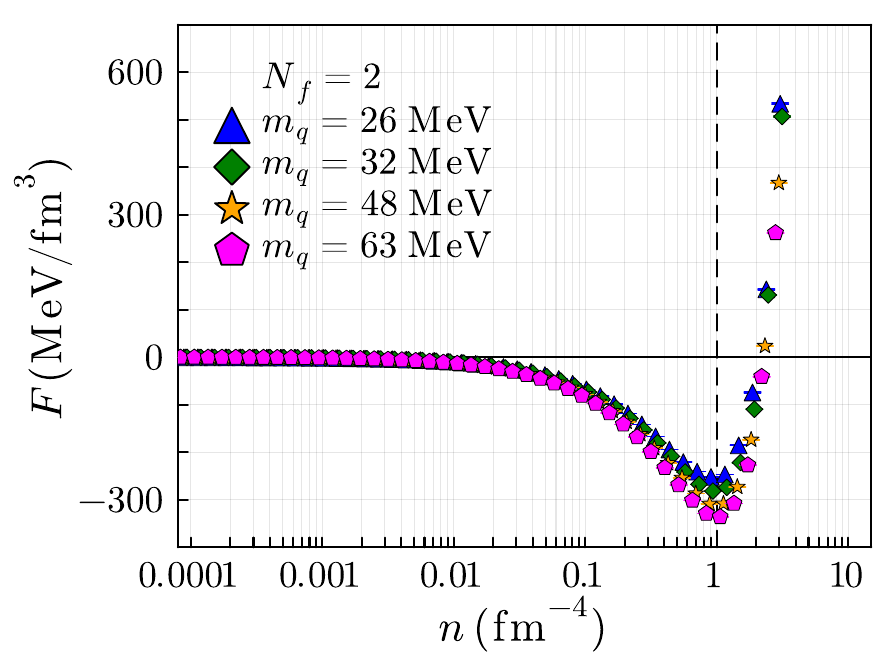}
  \caption{
    \label{fig:Nf2_F_vs_n} 
    Instanton density versus free energy 
    computed in IILM with 
    the flavor SU(2) symmetric quarks 
    of the current quark masses,
    $m_q=26,32,48$ and $63~{\rm MeV}$.
    }
\end{figure}

In Fig.~\ref{fig:Nf2_F_vs_n}, 
we show the free energy as a function of the instanton density
that is computed in the flavor SU(2) symmetric IILM. 
The free energy decreases as the instanton density 
increases and it increases beyond its minimum. 
This behavior clearly shows 
the attraction among instantons 
at the dilute density and 
the repulsive behavior in the dense density. 
An increase in the current quark mass 
reduces the minimum of the free energy. 
These attractive and repulsive interactions 
at the dilute and dense instanton density, 
and the reduction of the free energy 
are quantitatively agree with 
the results in the previous studies~\cite{Schafer:1996,Suda:2024}. 

\begin{figure}
  \includegraphics[width=0.46\textwidth]{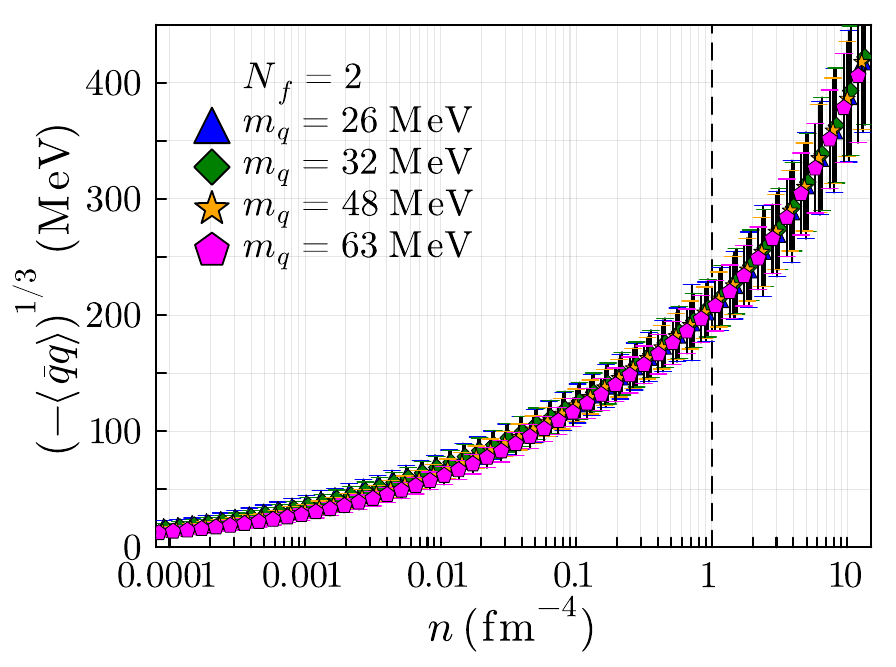}
  \caption{
    \label{fig:Nf2_qq_vs_n} 
    Instanton density versus cubic root of 
    light quark condensate in magnitude 
    in IILM with the flavor SU(2) 
    symmetric quarks. Their masses
    are equal to those in Fig.~\ref{fig:Nf2_F_vs_n}.
    }
\end{figure}

In Fig.~\ref{fig:Nf2_qq_vs_n}, we present 
the light quark condensate as a function of 
the instanton density. 
The absolute value of the quark condensate 
monotonically increase as the instanton density increases. 
At the vacuum of $n=1~{\rm fm}^{-4}$, the quark 
condensate takes about $\braket{\bar qq} = (-207~{\rm MeV})^3$
independent of the current quark mass. 
This value is consistent with the recent evaluation 
from lattice QCD calculation;
$\braket{0|\bar qq|0} = (\braket{0|\bar uu|0} + \braket{0|\bar dd|0})/2 = -[272(5)~{\rm MeV}]^3$~\cite{Gubler:2019}
even though our result is slightly underestimated.

Regarding the instanton density dependence of the quark condensate, 
we can compare our results with the theoretical 
predictions~\cite{Dyakonov:1986,Pobylitsa:1989,Wei-Yang:2025}.
Those predictions suggest that the quark condensate should be 
proportional to the square root of the instanton density in the 
large-$N_c$ limit at leading order in the instanton density 
expansion, and barely depends on the value of the current quark mass. 
For the sake of explanation, we denote an exponent of the instanton 
density $n$ by $r$, i.e., $\braket{\bar qq} \propto n^r$. From our 
results up to $n=0.001~{\rm fm^{-4}}$, this exponent takes 
$r=0.96, 0.99, 1.00, 1.00$ for $m_q=26, 32,48,63~{\rm MeV}$, respectively.
When we extend the range of data up to $n=0.5~{\rm fm^{-4}}$, the exponent 
reduces to $r=0.75, 0.77, 0.82, 0.85$, respectively. Although the 
exponent $r$ depends barely on the value of the current quark mass, 
we can find that the value of $r$ deviates from the theoretical prediction, 
$r=0.5$. This disagreement in the exponent could be understood as 
the effects of the finite current quark mass and the finite $N_c$.
According to Ref.~\cite{Wei-Yang:2025}, the quark condensate in the 
instanton density expansion starts from $n/m^*$ where $m^*$ is referred 
to as the determinantal mass given as a sum of the current quark mass $m$ 
and the contribution from the constituent quark mass. When there is the 
finite current quark mass, the quark condensate starts from the order 
of $\mathcal{O}(n)$. Thus, we consider that the present results are 
consistent with this theoretical prediction. 

\begin{figure}
  \includegraphics[width=0.46\textwidth]{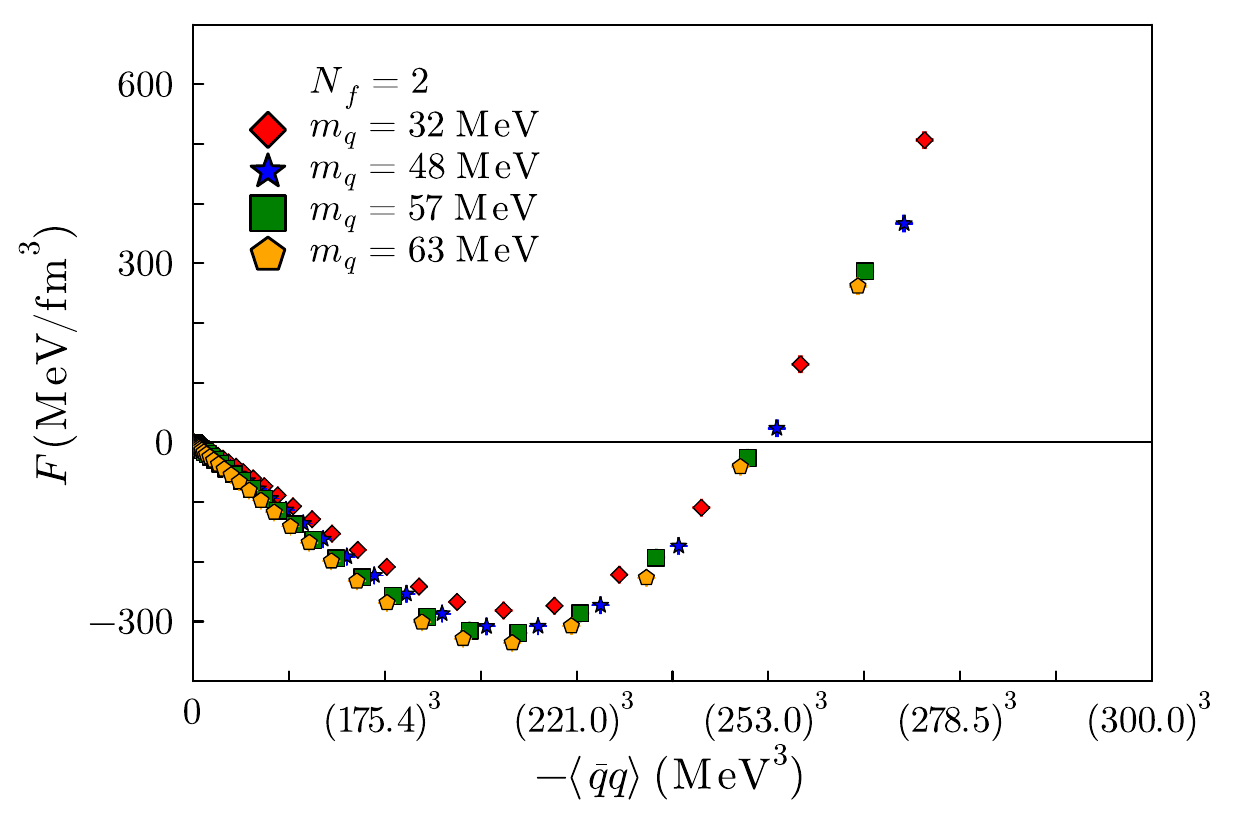}
  \caption{
    \label{fig:Nf2_F_vs_qq} 
    Scatter plot created using the values of the free energy $F$ 
    and the quark condensate $\braket{\bar{q}q}$ calculated 
    by the flavor SU(2) symmetric IILM at each density $n$, 
    with $\braket{\bar{q}q}$ on the horizontal axis 
    and $F$ on the vertical axis. 
    Different marker shapes represent different $m_q$ values.
  }
\end{figure}

In Fig.~\ref{fig:Nf2_F_vs_qq}, we show 
the plot of the dataset $(\braket{\bar qq}, F)$
computed by the flavor SU(2) symmetric IILM. 
The results correspond to parameter sets B3 through B6, 
with associated current quark masses 
summarized in Table~\ref{tabel:configuration_Nf2}. 
In all cases, 
the free energy exhibits a clear minimum 
at $\braket{\bar{q}q}\neq 0$. 
The finite value of the quark condensate 
provides direct evidence 
for the D$\chi$SB in the vacuum. 
The behavior of $F$ near 
$\braket{\bar{q}q}=0$ is approximately linear, 
reflecting the explicit breaking of chiral symmetry 
due to the finite current quark mass $m_q$.
This linear dependence is consistent 
with the result from the chiral effective theories 
in the small quark condensate region.

\begin{figure}
  \includegraphics[width=0.46\textwidth]{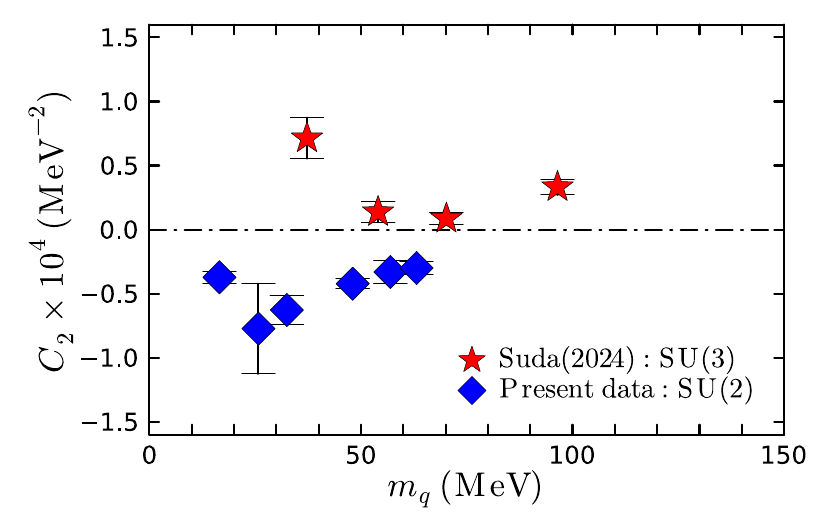}
  \caption{
    \label{fig:Nf3Nf2_C2_vs_mq} 
    Curvature $C_2$ evaluated in the flavor SU(2) symmetric IILM 
    for different quark masses are shown by blue diamonds, 
    which corresponding to Sets B1-B6. 
    The flavor SU(3) results are presented 
    with red stars from our previous study~\cite{Suda:2024}.
    }
\end{figure}

Performing the polynomial regression 
on the dataset $(\braket{\bar qq},F)$, 
we obtain the curvature at each quark mass $m_q$. 
For instance, we evaluate the curvature as $C_2 = -7.73\times 10^{-5}~{\rm MeV^{-2}}$ at 
the light quark mass of 
$m_q = 0.08~\Lambda = 26~{\rm MeV}$ by using the polynomial model of order $k=3$.
Figure~\ref{fig:Nf3Nf2_C2_vs_mq} 
shows the curvature evaluated 
at each current quark mass $m_q$ 
in the flavor SU(2) symmetric IILM. 
Across the entire range 
$17~\mathrm{MeV} < m_q < 63~\mathrm{MeV}$, 
we observe that $C_2$ remains negative, 
which contrasts with our previous results
obtained in the flavor SU(3) symmetric case~\cite{Suda:2024}, 
shown as red stars in Fig.~\ref{fig:Nf3Nf2_C2_vs_mq}
where $C_2$ is positive.

This difference in signs indicate
a qualitative change in the type 
of D$\chi$SB between 
the flavor SU(2) and SU(3) 
symmetric cases. 
The negative curvature implies that 
the ordinary type of D$\chi$SB occurs in the vacuum
for the flavor SU(2) symmetric case
according to our criterion. 

\subsection {The (2+1)-flavor calculation}
\label{subsec:2+1-flavor-res}

\begin{figure}
  \includegraphics[width=0.46\textwidth]{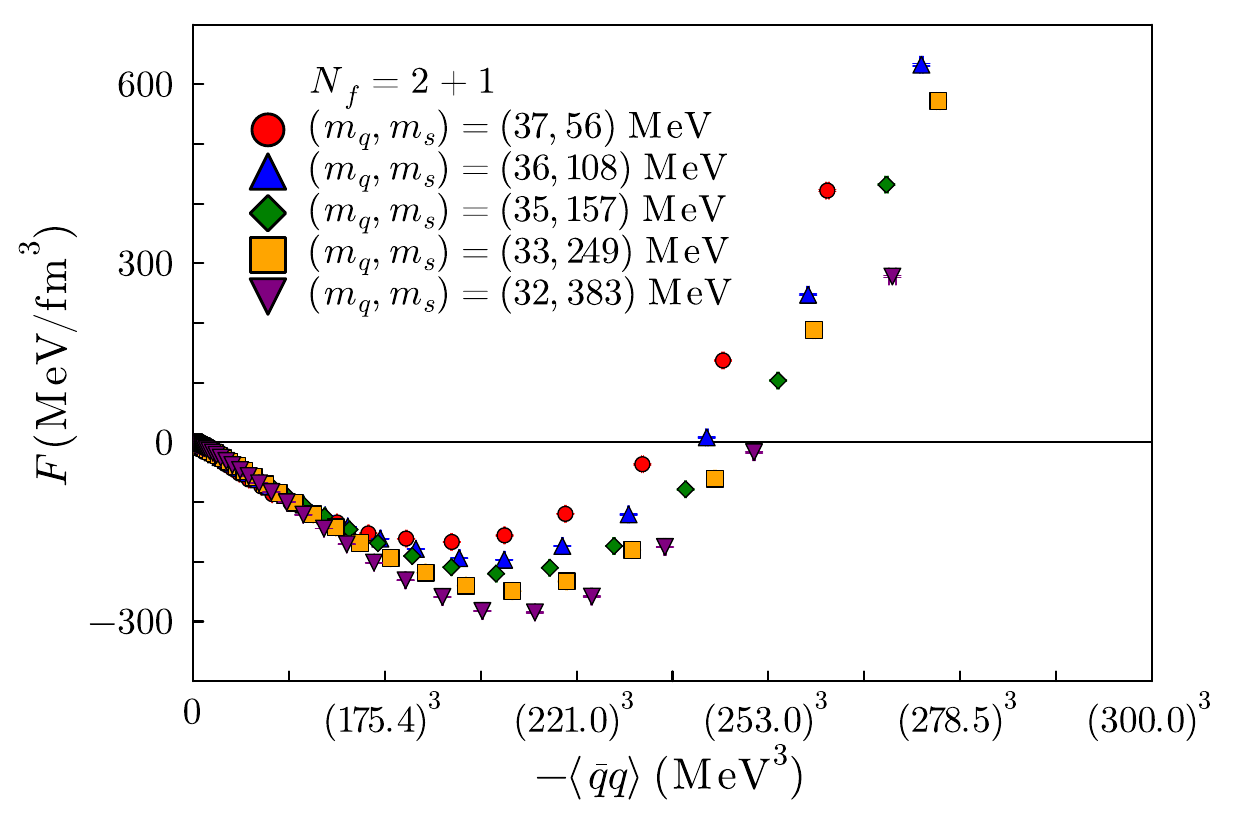}
  \caption{
    \label{fig:Nf21_F_vs_qq} 
    The free energy versus the quark condensate 
    computed in IILM with the (2+1)-flavor
    dynamical quarks masses,
    corresponding to Sets D1, D2, D3, D5 and D7. 
    }
\end{figure}

\begin{figure*}
  \includegraphics[width=0.92\textwidth]{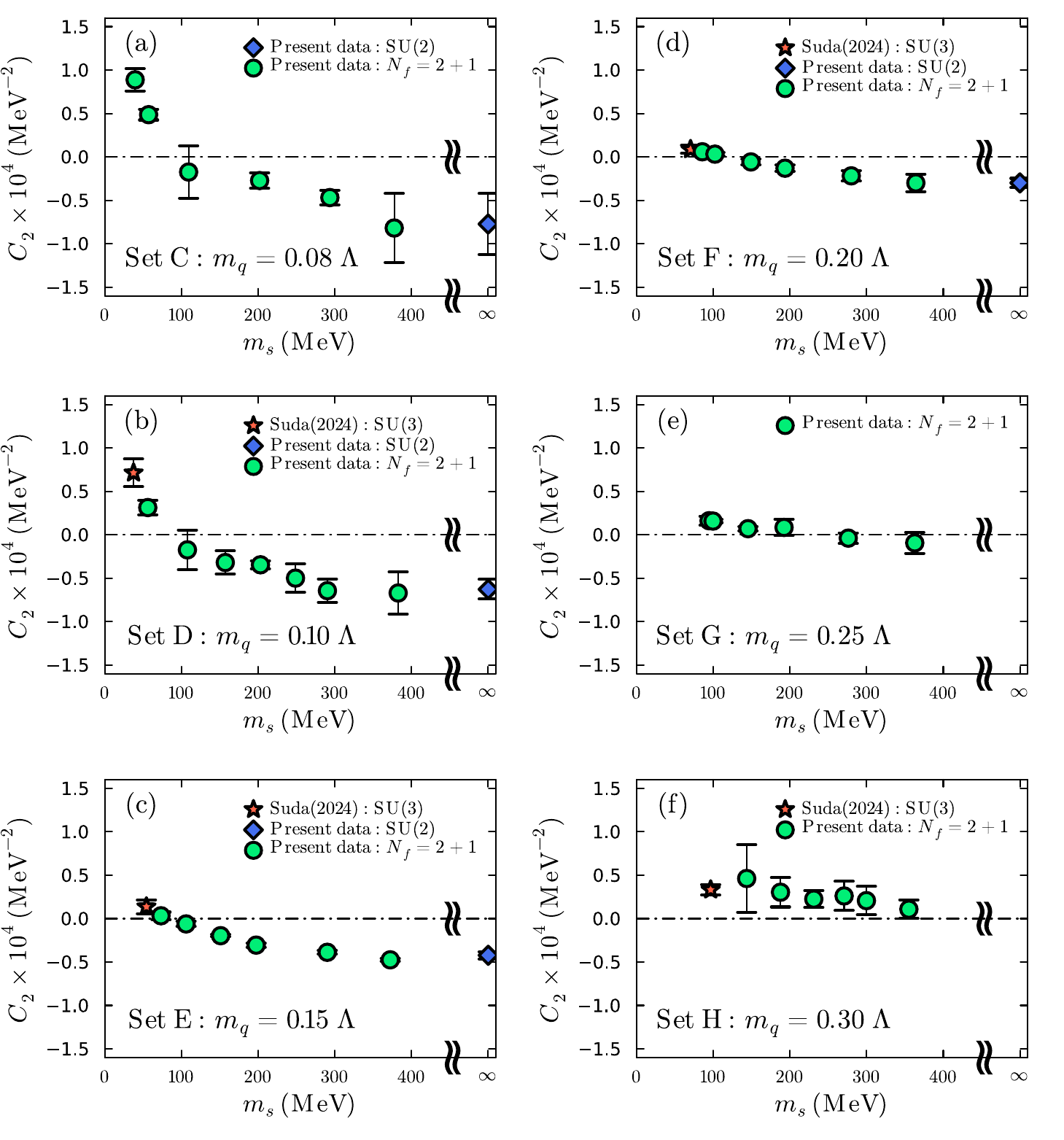}
  \caption{
    \label{fig:C2_Nf3_Nf2_Nf21_transition} 
    Plots of $m_s$ versus $C_2$ that 
    are evaluated by IILM with the (2+1)-flavor 
    quarks of Sets C-H. The values of $C_2$ are 
    obtained by polynomial regression of degree $k=3$.
    (a) Results of the flavor SU(2) symmetric 
    and the (2+1)-flavor calculations with 
    $m_q=0.08~\Lambda$ are shown by blue diamond 
    and green circle, respectively.
    (b), (c), (d) Results of the flavor SU(2) 
    symmetric, SU(3) symmetric, and the (2+1)-flavor 
    calculations with $m_q=0.10, 0.15$, and 
    $0.20~\Lambda$, respectively. Blue diamond, 
    orange star, and green circle denote each result. 
    (e) Results of the (2+1)-flavor calculations 
    with $m_q=0.25~\Lambda$. (f) Results of 
    the flavor SU(3) symmetric and the (2+1)-flavor 
    calculations with $m_q=0.30~\Lambda$ shown 
    by orange star and green circle, respectively. 
    The flavor SU(3) symmetric results are taken from 
    our previous study~\cite{Suda:2024}. 
    Results of the flavor SU(2) symmetric 
    calculation are presented on the point 
    of $m_s = \infty$.
    }
\end{figure*}

Figure~\ref{fig:Nf21_F_vs_qq} shows 
the dataset $(\braket{\bar qq},F)$ 
computed in IILM with the (2+1)-flavor quarks. 
In this plot, the light quark mass $m_q$ is fixed at $0.1~\Lambda$, 
while the strange quark mass $m_s$ is varied 
in the range $0.15~\Lambda \leq m_s \leq 1.2~\Lambda$ (see Table~\ref{tabel:configuration_Nf21}).

For all simulations at different quark masses, 
the free energy exhibits 
a minimum at a finite value of $\braket{\bar qq}$, 
indicating that D$\chi$SB occurs also in the vacuum 
of the (2+1)-flavor IILM. As $m_s$ increases, 
the free energy minimum becomes deeper and 
the magnitude of the quark condensate at the minimum becomes larger. 
This trend closely resembles the behavior observed 
in the flavor SU(3) IILM~\cite{Suda:2024}, 
suggesting that the strange quark mass 
slightly increases the absolute value of the 
quark condensate.

Figure~\ref{fig:C2_Nf3_Nf2_Nf21_transition} 
shows the curvature $C_2$ evaluated in IILM 
with the (2+1)-flavor quarks. 
Panels (a)-(e) correspond to different light quark masses $m_q$, 
and illustrate that $C_2$ is positive when the strange quark mass $m_s$ 
is smaller than $57$-$192~\mathrm{MeV}$, 
but becomes negative as $m_s$ exceeds these values.

These results suggest that IILM exhibits 
anomaly-driven type of D$\chi$SB 
in the nearly flavor SU(3) symmetric regime, 
characterized by positive curvature. 
In contrast, as the mass difference 
between the light and strange quarks increases, 
that is, as the system moves away from 
the flavor SU(3) symmetric limit, 
the system tends to favor 
the ordinary type of D$\chi$SB, 
indicated by negative curvature.

When the light quark mass is relatively 
heavy, such as $m_q = 0.3~\Lambda$ shown in panel (f), 
the curvature remains positive across 
the full range of $m_s$. 
This result is consistent with the observation
that the curvature is positive 
in the nearly flavor SU(3) 
symmetric case explained above.
In fact, in panel (f), the light quark mass is fixed 
at $m_q=0.3~\Lambda$, and the rightmost 
data point corresponds to the case with the heaviest 
$s$ quark mass, $m_s=1.2~\Lambda$. 
The mass ratio in this case is $m_s/m_q=4.0$. 
On the other hand, the third data point 
from the left in panel (a) corresponds to 
$m_q = 0.08~\Lambda$ and $m_s=0.30~\Lambda$, 
yielding a similar ratio $m_s/m_q=3.75$. 
From the viewpoint of the flavor SU(3) symmetry breaking 
characterized by the mass ratio $m_s/m_q$, 
the result in panel (f) should be compared with 
the left three data points in panel (a). 
In both cases, the curvature is positive within error,
indicating that the results are consistent. 
This observation motivates us to study the dependence 
of the curvature on the mass ratio $m_s/m_q$, 
addressed in the next section.

\subsection  {Comparison among the flavor SU(2), SU(3) symmetric and (2+1)-flavor calculations}
\label{subsec:transition-res}

In Fig.~\ref{fig:C2_Nf3_Nf2_Nf21_transition}, we have compared 
the curvature $C_2$ calculated in the (2+1)-flavor IILM with 
those obtained in the flavor SU(3) and SU(2) symmetric cases.
As the strange quark mass $m_s$ increases, the curvature smoothly 
decreases from a positive value that is characteristic of 
anomaly-driven type of D$\chi$SB to a negative value defining 
the ordinary type of D$\chi$SB. This can be seen in the change 
from the flavor SU(3) symmetric calculation to the flavor SU(2) 
symmetric case that is mediated by the (2+1)-flavor calculations.
This trend is clearly observed in the results for $m_q = 0.10$, $0.15$, 
and $0.20~\Lambda$, corresponding to panels (b), (c), and (d) of 
Fig.~\ref{fig:C2_Nf3_Nf2_Nf21_transition}. Although the degree 
of decrease becomes more gradual for heavier values of $m_q$, 
the qualitative behavior---namely, the decrease in curvature with 
increasing $m_s$---is consistently seen across all values of $m_q$.

These observations support our conjecture that IILM with nearly 
flavor SU(3) symmetric quark masses exhibits the anomaly-driven 
type of D$\chi$SB, whereas it tends to favor the ordinary type 
when the flavor SU(3) symmetry is explicitly broken by mass 
difference between the light and strange quarks.

\begin{figure}
  \includegraphics[width=0.46\textwidth]{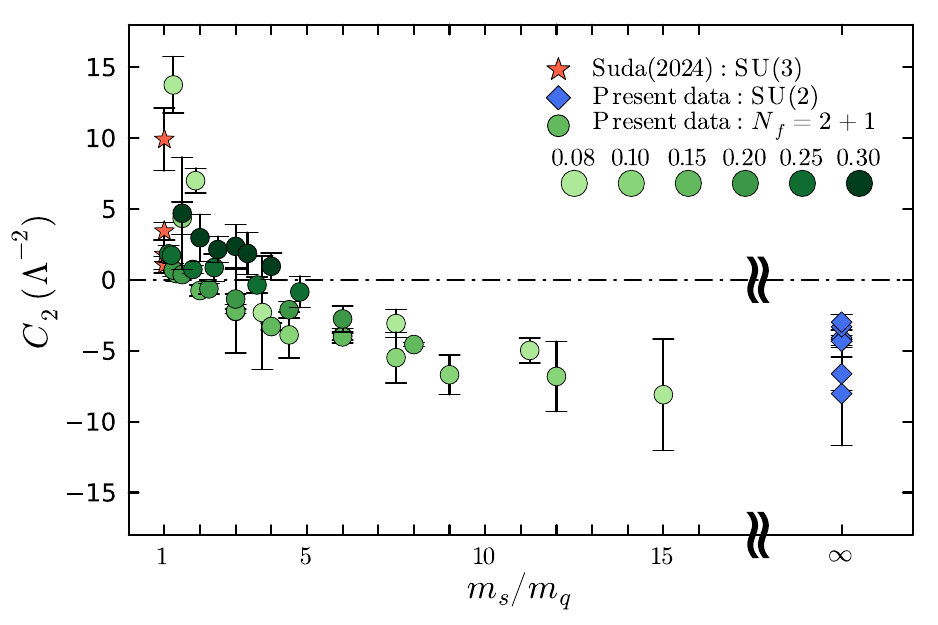}
  \caption{
    \label{fig:C2_mums_ratio} 
    $m_s$-to-$m_q$ ratio dependence of $C_2$. 
    The results of $C_2$ 
    in the flavor SU(3) and SU(2) symmetric 
    and the (2+1)-flavor calculations are presented 
    by red star, blue diamond and green circle, 
    respectively. For the (2+1)-flavor results, 
    the results with same $m_q$ in units of $\Lambda$ 
    are filled with same green gradation.
    }
\end{figure}

Figure~\ref{fig:C2_mums_ratio} shows the dependence of the curvature 
$C_2$ on the ratio of strange to light quark masses, $m_s/m_q$. The 
flavor SU(3) and SU(2) symmetric results correspond to $m_s/m_q = 1$ 
and $m_s/m_q \to \infty$, respectively, while the (2+1)-flavor calculations 
cover the intermediate region $1 < m_s/m_q \leq 15$ in the present study.

We clearly observe that $C_2$ decreases as the ratio increases, 
regardless of the absolute value of $m_q$. Figure.~\ref{fig:C2_mums_ratio} 
suggests that the curvature changes its sign from positive to 
negative around $m_s/m_q \sim 3$, and given this result, it is 
expected to become negative at the physical point $m_s/m_q = 27.3$.
In the limit of $m_s/m_q \to \infty$, the curvature eventually 
approaches negative value suggested by the results of the flavor 
SU(2) symmetric calculation.

This result suggests that the anomaly-driven type of D$\chi$SB 
is favored only within a narrow region where the flavor SU(3) 
symmetry is approximately emerged. Once this symmetry is broken 
by a mass difference among the $u$, $d$, and $s$ quarks, 
the ordinary type of D$\chi$SB emerges in the vacuum.
As we have mentioned in the end of Sect.~\ref{sec:Interactions}, 
the difference in number of quark flavors leads the different 
interaction between quarks via instantons~\cite{Shifman:1980}. 
If the system is in the flavor SU(3) symmetric regime, the quark 
determinant part in the interaction corresponds to the six-quark 
interaction, and thus this results the emergence of the anomaly-driven 
type of D$\chi$SB. While if one moves to the flavor SU(2) symmetric side, 
the heaviest quark flavor drops from the system and the quark determinant 
part only produces the four-quark interaction, leading the ordinary type 
of D$\chi$SB. Such a reduction of the determinant-type interaction to 
the four-quark one in two-flavor system has also been discussed in the 
context of the NJL-based analysis~\cite{Bentz:1999,Itakura:2000,Liu:2023,Liu:2024}. 
We observe this in the language of the instanton liquid model. 

\begin{figure}
  \includegraphics[width=0.46\textwidth]{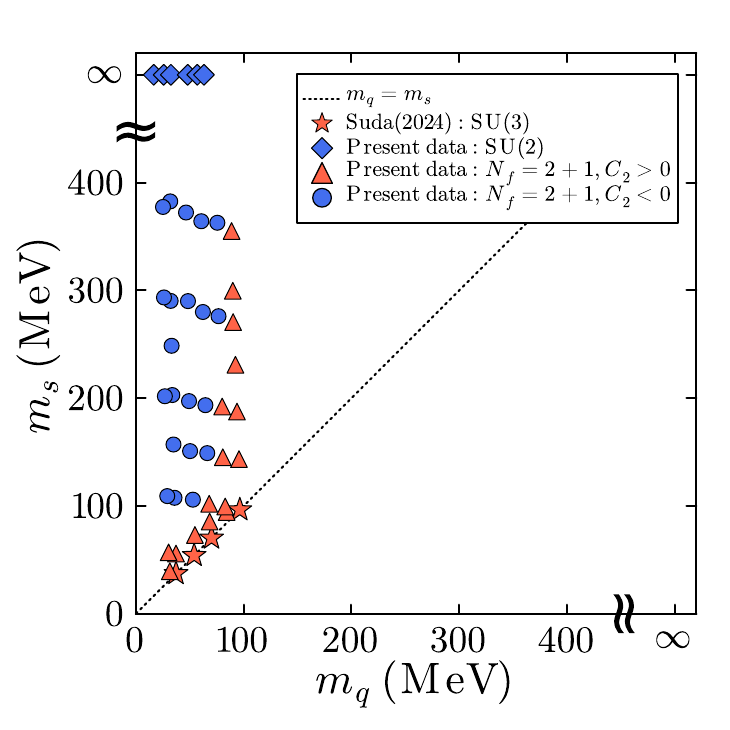}
  \caption{
    \label{fig:mud_ms_phase_diagram} 
    Our simulation points in the plane of $m_q$ and $m_s$.
    Points corresponding to the flavor SU(2) 
    symmetric calculations are symbolically 
    located on the axis $m_s=\infty$. 
    }
\end{figure}

Figure~\ref{fig:mud_ms_phase_diagram} summarizes the distribution 
of the simulations at different quark masses computed in this study 
in the plane of $m_q$ and $m_s$. In this figure, each marker corresponds 
to specific quark masses: the flavor SU(3) symmetric calculations are 
plotted with star, the flavor SU(2) symmetric calculations are denoted 
by diamond, and the (2+1)-flavor calculations with a positive and a 
negative curvature are shown with triangle and circle, respectively. 
The flavor SU(3) symmetric and a part of the (2+1)-flavor calculations 
with triangle marker have a positive curvature, associated with the 
anomaly-driven type of D$\chi$SB. While the flavor SU(2) symmetry and 
a part of the (2+1)-flavor calculations with circle marker have a 
negative curvature, corresponding to the ordinary type of D$\chi$SB.

In the region where $m_q < 80~\mathrm{MeV}$, we observe that the markers 
with a positive curvature are clustered just above the flavor SU(3) 
symmetric line, $m_q = m_s$. This suggests that the anomaly-driven type 
of D$\chi$SB can still occur even when the flavor SU(3) symmetry is 
slightly broken. However, as $m_s$ increases to around $80~\mathrm{MeV}$ 
with keeping $m_q$ fixed below this value, the markers with a negative 
curvature appear, indicating a change in the type of D$\chi$SB from 
the anomaly-driven to the ordinary. This result further supports our 
expectation that the anomaly-driven type of D$\chi$SB is emerged only 
in the vicinity of the flavor SU(3) symmetry.

\section  {Conclusions}
\label{sec:conclusion}

In this study, we have investigated the type of dynamical 
chiral symmetry breaking (D$\chi$SB) using the interacting 
instanton liquid model (IILM) with the flavor SU(2) symmetric 
and (2+1)-flavor quarks. Following our previous work~\cite{Suda:2024}, 
we defined the type of D$\chi$SB based on the sign of the 
second derivative of the free energy with respect to the quark 
condensate at the origin.

Our calculations show that in the flavor SU(2) symmetric case 
the curvature is negative. A negative curvature means the emergence 
of the ordinary type of D$\chi$SB by our criterion. This contrasts 
with the results of a positive curvature obtained in the flavor SU(3) 
symmetric calculation~\cite{Suda:2024}. 

In the (2+1)-flavor calculations, we have found that the curvature decreases 
as the strange quark mass increases and eventually becomes negative. As we 
increase the strange quark mass, the (2+1)-flavor calculation effectively 
changes from the flavor SU(3) symmetric calculation to the flavor SU(2) 
symmetric one. In this change, we have observed that when the strange quark 
mass exceeds approximately three times those of the light quarks, the 
curvature turns negative. This implies that the anomaly-driven type of 
D$\chi$SB, characterized by a positive curvature, emerges for a small ratio 
of $m_s/m_q$, while the ordinary type of D$\chi$SB occurs with a large ratio. 
This also suggests that the magnitude of the strange quark mass plays a crucial 
role in determining the predominant chiral symmetry breaking mechanism. 

It is interesting direction to investigate how the competition 
between the flavor SU(3) symmetry breaking and the effects of the \UA
anomaly emerges in the hadron spectrum using the chiral effective theories 
that includes the 't Hooft vertex~\cite{Kuroda:2020,Saionji:2023}.

The present results may be relevant to discussions regarding 
the nature of the chiral phase transition at finite temperature. 
In an earlier instanton-based theoretical analysis, the authors 
of Ref.~\cite{Schafer:1996} concluded that in the system with two 
light and one intermediate mass quark flavor, the chiral phase 
transition occurs at $T \simeq 140~{\rm MeV}$, although it remains 
inconclusive whether the transition is first or second order. They 
also emphasized the importance of the finite-size effects in the 
determination of the critical temperature. Recent studies shed 
light on the significance of the number of quark flavors and 
the role of the axial anomaly in determining the order of chiral 
phase transition~\cite{Pisarski:2024, Fejos:2025, Giacosa:2025}.
Since the current work focuses on the mechanism responsible for 
chiral symmetry breaking already emerges at zero temperature, 
a detailed investigation of its connection to finite-temperature 
phenomena is left for future studies.

In order to obtain the clue to the understanding of 
the QCD vacuum structure, we desire information about 
the regions where the quark masses are lighter than 
present values. To make the quark mass light, we need 
to enlarge the simulation box size while keeping the 
instanton density fixed to ensure the validity of 
treating the light quarks. Achieving this requires 
the increasing the total number of instantons and 
anti-instantons in the simulation, which in turn 
demands more computational resources. This will be 
addressed in future work. 

%
%
\section*{Acknowledgments}
\noindent
This work of Y.~S. was supported 
by JST SPRING, Japan (JPMJSP2106 and JPMJSP2180), 
and by a Grant-in-Aid for JSPS Fellows (JP25KJ1278).
The work of D.J. was supported in part 
by Grants-in-Aid for Scientific Research from 
JSPS (JP21K03530, JP22H04917, JP23K03427 and JP25K07315).
%
%
%

%
%
%

\appendix 

\section  {Semiclassical instanton amplitude}
\label{app:semiclassicalamp}
In this study, we use the semiclassical instanton amplitude 
$f(\rho)$ with up to two-loop $\beta$ functions~\cite{tHooft:1976_b}. 
The explicit form is given by 
\begin{eqnarray}
  f(\rho) 
  &=& C_{N_c} \left[ \frac{8 \pi^2}{g^2(\rho)} \right]^{2N_c} 
  \exp \left[ - \frac{8\pi^2}{g^2(\rho)} \right] 
  \frac{1}{\rho^5} \nonumber \\ 
  &=& C_{N_c} \frac{1}{\rho^5} \beta_1(\rho)^{2N_c} \nonumber \\ 
  &\ & \times \exp \left[ -\beta_2(\rho) + \left( 2N_c - \frac{b'}{2b} \right)
  \frac{b'}{2b} \frac{1}{\beta_1(\rho)} \ln \beta_1(\rho) \right], \nonumber \\  \\
  C_{N_c} &=& \frac{0.466 e^{-1.679 N_c}}{(N_c-1)! (N_c-2)!}, 
\end{eqnarray}
where the gauge coupling $g^2$ is given as a function of the instanton size $\rho$; 
the $\beta$ functions $\beta_1,\beta_2$ include up to two-loop order; and the Gell-Mann--Low coefficients 
are given as follows: 
\begin{eqnarray}
  \beta_1(\rho) &=& -b \ln(\rho \Lambda), \quad 
  \beta_2(\rho) = \beta_1 (\rho) + \frac{b'}{2b} \ln \left[ \frac{2}{b} \beta_1 (\rho) \right], \nonumber \\ \\
  b &=& \frac{11}{3}N_c - \frac{2}{3} N_f, \quad 
  b' = \frac{34}{3} N_c^2 - \frac{13}{3} N_c N_f + \frac{N_f}{N_c}. \nonumber \\
\end{eqnarray}
Here, the number of colors and flavors are denoted as $N_c$ and $N_f$. 
$\Lambda$ is the scala parameter described in the text.

\section  {Explicit expression of the interactions}
\label{app:interaction}

The explicit expression of $S_{\rm int}^{(2)}(l,m)$ 
is given by 
\begin{eqnarray}
  S^{(2)}_{\rm int}(l,m) = 
  \begin{cases}
    S^{(2)}_{IA} + S^{(2)}_{\rm core} & {\rm for\ } (l,m) = (I,A)\ {\rm or\ } (A,I) \\
    S^{(2)}_{\rm core} & {\rm for\ } (l,m) = (I,I)\ {\rm or\ } (A,A)
  \end{cases}, \nonumber \\
\end{eqnarray}
where $I,A$ denote the instanton and anti-instanton, respectively. 
$S^{(2)}_{IA}$ is the two-body action that 
is calculated by using the streamline configuration~\cite{Verbaarschot:1991}, 
and $S^{(2)}_{\rm core}$ represents the phenomenological repulsive core 
at the short distance. Their explicit forms are derived in Ref.~\cite{Schafer:1996} as 
\begin{eqnarray}
  \frac{S_{IA}^{(2)}}{S_0}
  &=& \frac{1}{(\lambda^2 - 1)^3} \left\{ 
    -4 [1-\lambda^4 + 4 \lambda^2 \ln (\lambda)] (|u|^2 - 4 |u \cdot \hat R|^2) \right. \nonumber \\ 
    &\ & \left. 
    + 2[1-\lambda^2 + (1+\lambda^2) \ln (\lambda) ] \right. \nonumber \\ 
    &\ & \left. 
    \times 
    [ (|u|^2 - 4|u \cdot \hat R|^2)^2 + |u|^4 + 2(u)^2 (u^*)^2 ]
  \right\}, \\
  \frac{S_{\rm core}^{(2)}}{S_0} 
  &=& \frac{A}{\lambda^4}|u|^2, 
\end{eqnarray}
where the single instanton action is calculated as $S_0=8\pi^2 / g^2(\rho) = \beta_1(\bar \rho)$ 
with $\bar \rho \equiv \sqrt{\rho_I \rho_A}$. 
The parameter $\lambda$, called as the conformal parameter, 
is introduced in construction of the streamline ansatz~\cite{Yung:1988}, given by 
\begin{eqnarray}
  \lambda 
  &=& \frac{1}{2} \frac{R^2 + \rho_l^2 + \rho_m^2}{\rho_l \rho_m} + \frac{1}{2} \sqrt{\frac{(R^2 + \rho_l^2 + \rho_m^2)^2}{\rho_l^2 \rho_m^2} - 4}, \nonumber \\
\end{eqnarray}
with $R_\mu = z_{l,\mu} - z_{m,\mu}$ and $R^2 = R_\mu R_\mu\ (\mu=1,2,3,4)$. 
$\hat R_\mu$ is a unit vector of $R_\mu$.
The four-vector $u_\mu$ characterizes the relative color orientation between 
instantons of color degrees of freedom $U_l$ and $U_m$,
given by 
\begin{eqnarray}
  u_\mu = \frac{1}{2i} \Tr (U_l^\dag U_m \tau_\mu^+), 
\end{eqnarray}
with the four dimensional Pauli matrix $\tau^\pm_\mu = (\tau_i, \mp i \bm{1}_2)$. 
With those vectors, the inner product $(u \cdot \hat R)$ is understood as 
a contraction: $(u \cdot \hat R) = \sum_{\mu=1}^4 u_\mu \hat R_\mu$. 

The overlap matrix $T$ has the structure of 
\begin{eqnarray}
  T = 
  \begin{pmatrix}
    \bm{0}_{N_+ \times N_+} & (\mathcal{T})_{N_+ \times N_-} \\ 
    (\mathcal{T})_{N_- \times N_+} & \bm{0}_{N_- \times N_-} \\ 
  \end{pmatrix},
\end{eqnarray}
with the matrix element
\begin{eqnarray}
  (\mathcal{T})_{IJ} = \int d^4x \psi^{*}_{0,I}(x;U,\rho,z) i \gamma_\mu D_\mu \psi_{0,J}(x;U,\rho,z). \qquad 
  \label{eq:T_def}
\end{eqnarray}
Since we only consider the system with $N_+=N_-$, the overlap matrix is square. 
The explicit expression of matrix element in Eq.~(\ref{eq:T_def}) 
is given in Ref.~\cite{Verbaarschot:1992,Schafer:1996} as 
\begin{eqnarray}
  (\mathcal{T})_{IJ} &=& i (u \cdot \hat R) \frac{1}{\sqrt{\rho_I \rho_J}} F(\lambda), \\ 
  F(\lambda) &=& 6 \int_0^\infty \frac{r^{3/2}}{(r+1/\lambda)^{3/2}(r + \lambda)^{5/2}} dr \nonumber \\ 
  &\approx& \frac{c_2\lambda^{3/2}}{[1+1.25 (\lambda^2 - 1) + c_2 (\lambda^2-1)^2]^{3/4}},
\end{eqnarray}
with $c_1=3\pi/8$ and $c_2 = (3\pi/32)^{4/3}$.

%
%

\end  {document}